\documentclass[conference]{IEEEtran}
\IEEEoverridecommandlockouts
\usepackage{amsmath,amssymb,amsfonts}
\usepackage{array,multirow,color,booktabs,textcomp}
\usepackage{algorithmic}
\usepackage{float}
\usepackage{hyperref}
\usepackage{mathtools}
\usepackage{bm}
\usepackage{amssymb}
\usepackage{preview}
\usepackage{enumitem}
\usepackage{caption}
\usepackage{balance}
\usepackage{threeparttable,booktabs}
\usepackage{tipa}
\usepackage{algorithm}
\usepackage{algorithmic}
\usepackage{multirow}
\usepackage{subcaption}
\usepackage{cite}
\usepackage[table]{xcolor}
\newcolumntype{g}{>{\columncolor{Gray}}c}

\usepackage{comment}
\usepackage{tikz}
\usepackage{amsmath}
\usepackage{arydshln}
\def\BibTeX{{\rm B\kern-.05em{\sc i\kern-.025em b}\kern-.08em
    T\kern-.1667em\lower.7ex\hbox{E}\kern-.125emX}}

\begin{document}

\title{Mouth Articulation-Based Anchoring for Improved Cross-Corpus Speech Emotion Recognition\\
%\title{Enhancing Cross-Corpus Speech Emotion Recognition Through Articulatory Gesture Constraints\\
}
\author{Shreya G. Upadhyay${^1}$, Ali N. Salman${^2}$,  Carlos Busso$^{2,3}$, Chi-Chun Lee${^1}$   \\
    ${^1}$Department of Electrical Engineering, National Tsing Hua University, Taiwan. \\
    ${^2}$Department of Electrical and Computer Engineering, University of Texas at Dallas, USA.\\
    ${^3}$Language Technologies Institute, Carnegie Mellon University, USA\\
    shreya@gapp.nthu.edu.tw, ali.salman@utdallas.edu, busso@cmu.edu, cclee@ee.nthu.edu.tw
}

\maketitle

\begin{abstract}
Cross-corpus \emph{speech emotion recognition} (SER) plays a vital role in numerous practical applications. Traditional approaches to cross-corpus emotion transfer often concentrate on adapting acoustic features to align with different corpora, domains, or labels. However, acoustic features are inherently variable and error-prone due to factors like speaker differences, domain shifts, and recording conditions. To address these challenges, this study adopts a novel contrastive approach by focusing on emotion-specific \emph{articulatory gestures} as the core elements for analysis. By shifting the emphasis on the more stable and consistent articulatory gestures, we aim to enhance emotion transfer learning in SER tasks. Our research leverages the CREMA-D and MSP-IMPROV corpora as benchmarks and it reveals valuable insights into the commonality and reliability of these articulatory gestures. The findings highlight mouth articulatory gesture potential as a better constraint for improving emotion recognition across different settings or domains.
\end{abstract}

\begin{IEEEkeywords}
speech emotion recognition, articulatory gestures, cross-corpus, transfer learning.
\end{IEEEkeywords}

\section{Introduction}
\emph{Speech emotion recognition} (SER) systems are essential for improved user experiences across various applications, including automated call centers, education, entertainment, and medical fields \cite{narayanan2013behavioral, acosta2009using, tawari2010speech, devillers2010real}. In cross-corpus SER, aligning corpora from different domains and settings poses a significant challenge \cite{zhang2021deep}. Existing research has introduced several techniques to address domain, label, and feature discrepancies, such as transfer learning, semi-supervised learning, and few-shot or zero-shot learning to improve model generalization \cite{parthasarathy2020semi, ahn2021cross, xu2023zero}. Numerous approaches like optimizing distance metrics \cite{Gideon_2021}, adversarial training \cite{Abdelwahab_2018_3}, GANs to generate synthetic data \cite{su2021conditional}, and phonetic-based feature alignments \cite{upadhyay2023an} have also been explored. These SER methods mostly focused on handling acoustic feature mismatch between corpora, given their strong correlation with emotion and ease of recording. 

In our previous work \cite{upadhyay2023phonetic}, we introduced a phoneme-anchoring approach to enhance cross-corpus alignment in SER. This method focused on identifying stable sub-units by leveraging shared vowel-phoneme emotion-specific acoustic spaces to match acoustic distributions across corpora. By establishing stable phoneme-based anchors, we hypothesized that similar phonemes would yield similar acoustic features. Unlike many previous approaches that attempted to directly match acoustic feature distributions, our stable phoneme-anchoring method led to improved SER performance in cross-corpus settings. However, a critical question remains: Are acoustic features the most stable anchors for cross-corpus alignment in SER tasks? While acoustic features play a key role in conveying emotion, they are also vulnerable to noise, microphone quality, and recording environment variations, which can undermine SER accuracy in cross-corpus scenarios. We believe that our previous idea can achieve further improvement by incorporating more stable anchoring units.

Acoustic signals and articulatory features are intrinsically linked, offering complementary insights into emotion recognition \cite{el2011survey, li2024speech, zhang2023study}. Significant research has explored mapping between these modalities, such as converting acoustic to articulatory and vice versa \cite{li2024speech}. Emotions are closely tied to articulatory movements, particularly in facial expressions, where the mouth region is crucial due to its role in speech production \cite{kim2014say}. Unlike broader facial features, mouth gesture are more stable because of their limited physical range, making them valuable for emotion recognition tasks \cite{sadoughi2018expressive, shah2019articulation}. Focusing on the more stable aspect of speech production (such as \emph{articulatory gesture}) can offer a promising alternative. In this study, we adopt the definition of \emph{articulatory gestures} (AG) as the coordinated actions of speech organs (in this case, the mouth) that produce distinct phonemes. 
%Unlike variable and error-prone acoustic features, AG which are driven by muscle actions, are less influenced by external factors.
%In this study, by conditioning on these stable AG properties, we aim to improve cross-corpus alignment which can offer a more reliable foundation for transfer learning in SER under cross-corpus settings. 
This study aims to improve cross-corpus alignment using stable AG properties, hypothesizing that stable mouth articulation patterns should result in similar acoustic characteristics.

Mouth articulation data is available through methods like electromagnetic articulography (EMA) \cite{erickson2016articulation} and real-time magnetic resonance imaging (MRI) \cite{kim2020vocal, narayanan2004approach}. However, these methods are challenging to record and have limited corpora. Inspired by past research using marker information to analyze AG \cite{wang2013articulatory, sadoughi2018expressive}, this work focuses on using mouth landmarks extracted from the visual modality as a representation of AG. This work introduces the concept of incorporating constraints on AG into transfer learning to improve emotion recognition accuracy across corpora. We evaluate our approach using two multimodal datasets, CREMA-D \cite{cao2014crema} and MSP-IMPROV \cite{busso2016msp}. Our proposed cross-modal anchoring idea, \emph{articulatory gesture-anchored cross-corpus SER} (AG-CC), shows improved performance compared to the considered baseline.

\section{ Articulatory Gesture Analysis}
\label{AG_analysis}

\subsection{Multi-Modal Affective Corpora}

\smallskip
\noindent
The \textbf{CREMA-D} \cite{cao2014crema} (CREMA) is a publicly available resource for emotion recognition research. It includes approximately 7.5 hours of recordings from 91 actors, each performing seven categorical emotions, and primary attributes across 12 scripts. With around 7,440 utterances averaging 3 to 4 seconds each, the corpus provides a rich source of multimodal emotional expressions through both audio and video recordings.

\smallskip
\noindent
The \textbf{MSP-IMPROV} \cite{busso2016msp} (IMPROV) corpus includes approximately 8.5 hours of recordings, consisting of 8,438 prompted and spontaneous emotional sentences, with each utterance averaging about 4 seconds in length. This corpus is specifically designed for emotion recognition tasks and provides both audio and video recordings.  
%totaling 5,673 utterances. 
%We leverage the visual information from the corpus to extract facial landmarks, which facilitates the analysis of articulatory gestures. 

\smallskip
\noindent
We select these corpora for their diverse emotional expressions and multimodal data. IMPROV provides naturalistic, conversational emotions with varied intensity, while CREMA offers more controlled, scripted emotional expressions that are often more intense.
In this study, we only focus on four major emotions: \emph{Neutral, Anger, Happiness}, and \emph{Sadness}. 
%The visual modality is used to extract mouth landmarks, which becomes base for our mouth articulatory gesture analyses. 
The phoneme information for both the corpora is obtained using the Montreal Forced Aligner (MFA) \cite{mcauliffe2017montreal}.

\subsection{AG Feature Extraction and Prepossessing}
Our goal in this step is to extract robust features that enable the comparison of the mouth region across different subjects and corpora. To achieve this, we first use OpenFace \cite{baltruvsaitis2016openface} to detect the face bounding box and identify 68 2D landmark points that capture consistent facial features (e.g., eyes, chin, lips). We then rotate the landmarks to align the distance between the eyes parallel to the x-axis and normalize by the inter-pupil distance, reducing speaker-specific variations and ensuring consistent results from the same speaker across different sessions \cite{Ren_2016, Feng_2018}. This work examines mouth region AG by analyzing twelve key landmarks (48 to 59) that define the outer mouth shape. Building on insights from our previous research \cite{upadhyay2023phonetic}, we focus on six vowel phonemes: \{\textipa{A}, \textipa{@}, \textipa{E}, \textipa{i}, \textipa{\ae}, \textipa{u}\}. Phoneme-specific AG segments are extracted by segmenting frames based on phonetic boundaries. %To ensure clean training and testing, only the validation sets from corpora are used.

\begin{figure}[t!]
\centering
    \begin{subfigure}{0.5\textwidth}
       \includegraphics[width=\textwidth]{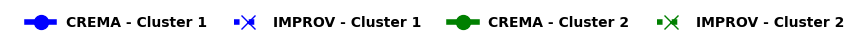}
    \end{subfigure}
    
    \begin{subfigure}{0.24\textwidth}
       \includegraphics[width=\textwidth]{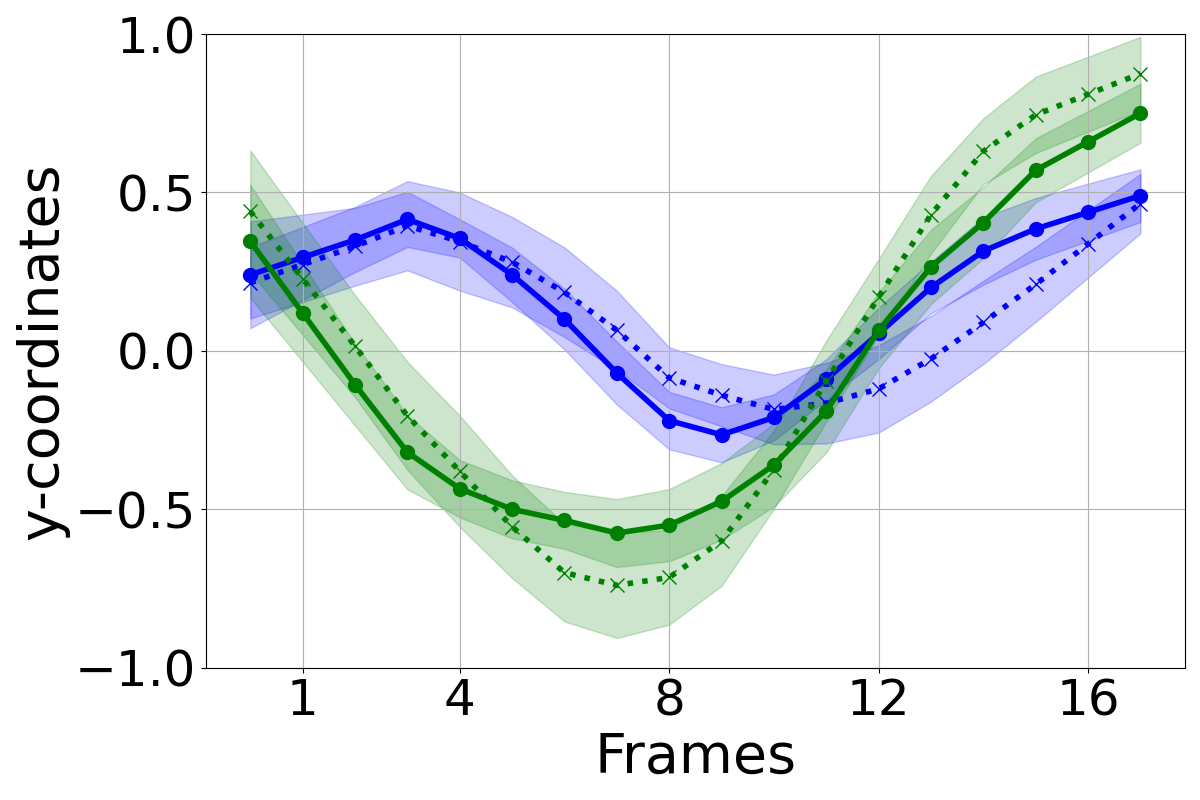}
       \subcaption{\emph{/\textipa{A}/  (Lower-Mid Point)}}
       \label{fig:aa}
    \end{subfigure}
    \begin{subfigure}{0.24\textwidth}
       \includegraphics[width=\textwidth]{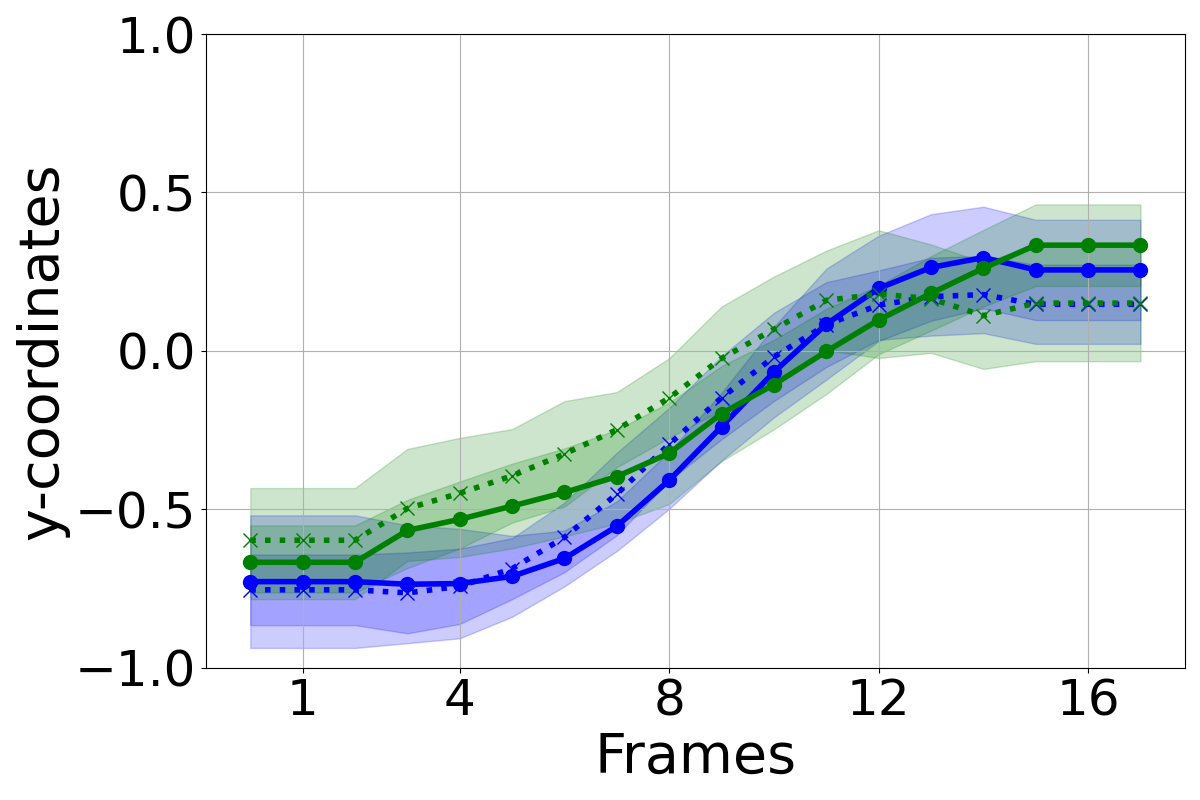}
       \subcaption{\emph{/\textipa{i}/ (Lower-Mid Point)}}
       \label{fig:i}
    \end{subfigure}  
    
    \begin{subfigure}{0.24\textwidth}
       \includegraphics[width=\textwidth]{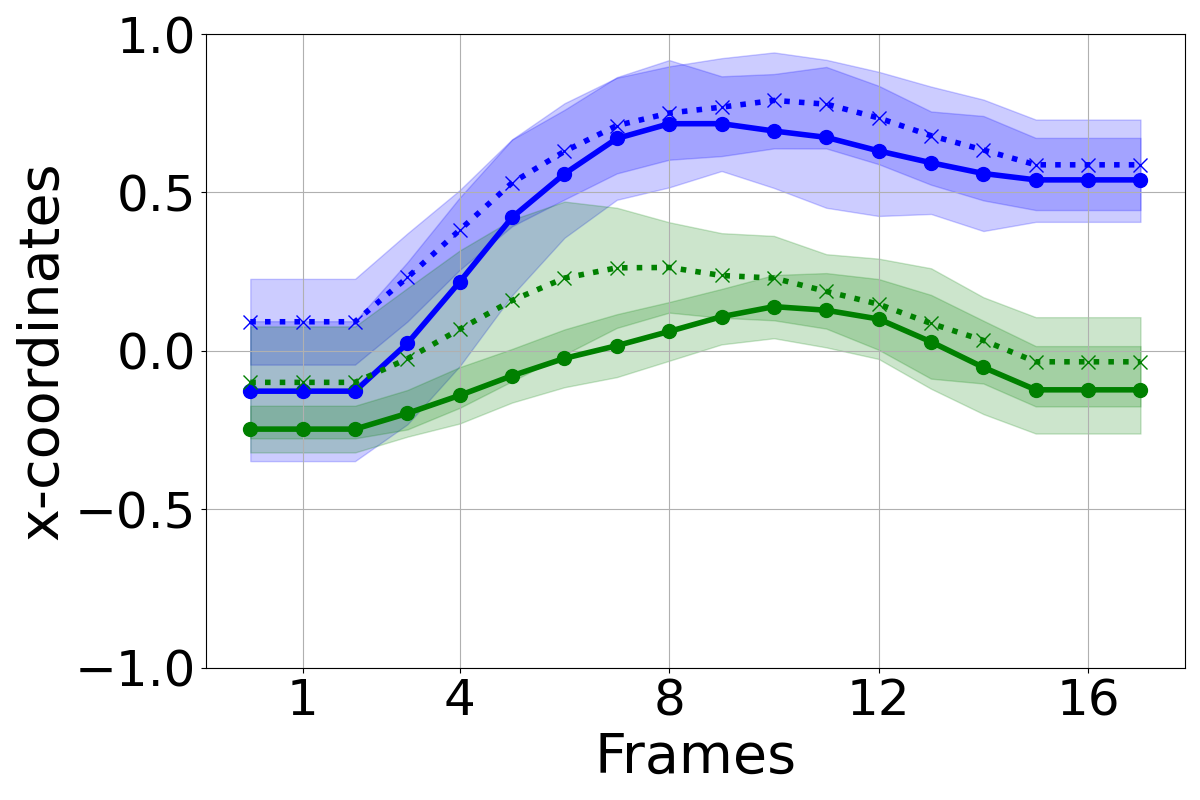}
       \subcaption{\emph{/\textipa{i}/ (Right-Corner Point)}}
       \label{fig:i_c1}
    \end{subfigure}
    \begin{subfigure}{0.24\textwidth}
       \includegraphics[width=\textwidth]{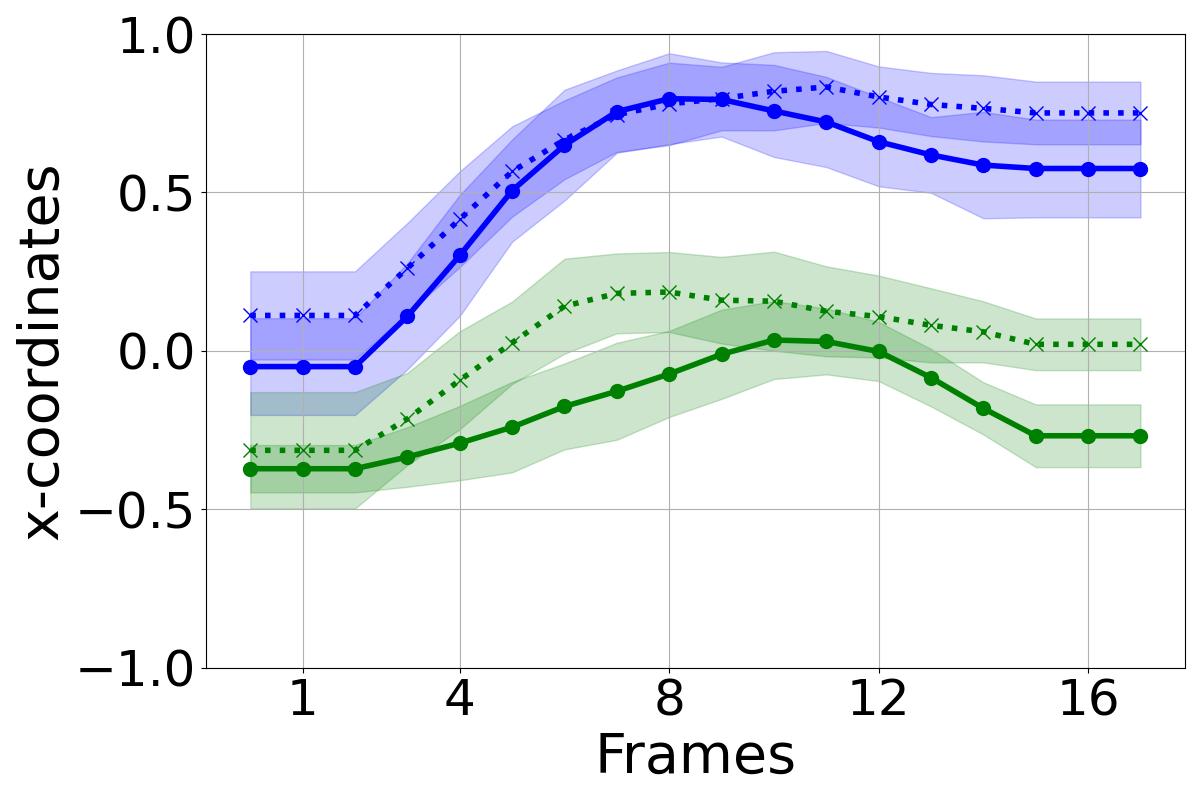}
       \subcaption{\emph{/\textipa{i}/ (Left-Corner Point)}}
       \label{fig:i_c2}
    \end{subfigure}  
\caption{Clustered AG patterns for /\textipa{A}/ and /\textipa{i}/ from both corpora come from two AG clusters; shows the mean pattern at each frame, with standard deviation indicated over 50 samples for each vowel.}
\label{fig:clusters}
\end{figure}
\subsection{Articulatory-Gesture Clustering} 
\label{sec:clustering}

Mouth gestures are continuous and dynamic, making hard segmentation difficult without clear boundaries. Clustering provides a more effective approach for capturing distinct AG patterns. First, we segment the long landmark sequences into smaller, contextually relevant segments using phonetic boundaries. We then apply time-series \textit{k}-means clustering \cite{huang2016time} with Soft-DTW (Dynamic Time Warping) as the distance metric, grouping mouth shapes with similar articulatory patterns despite timing variations. Validation samples from all four emotion categories across both corpora are used to train the AG cluster model. Testing different \textit{k} values (5 to 30), optimal cluster number is found to be 10 using the elbow method.

To evaluate the clustering model, we analyze how AG pattern variations for the same vowel are captured across different clusters. Fig. \ref{fig:clusters} presents examples using the vowels /\textipa{A}/ and /\textipa{i}/. For this analysis, we focus on key points: the corner-left (48th landmark) and corner-right (54th landmark) for x-coordinate analysis and the lower mid-mouth point (57th landmark) for y-coordinate analysis.
Fig. \ref{fig:aa} and Fig. \ref{fig:i} show y-coordinate curves which is aligned with expected mouth movements. For /\textipa{A}/, the downward AG curve in Fig. \ref{fig:aa} reflects the lower mid-point moving down, while the upward trend in Fig. \ref{fig:i} for /\textipa{i}/ shows the lower mid-point rising, both consistent with expected articulation. For the x-coordinates, examining the corner points in Fig. \ref{fig:i_c1} and Fig. \ref{fig:i_c2} aligns with expected behavior, as the x-coordinates increase during the pronunciation of /\textipa{i}/ in their respective direction.
The plots shown in Fig. \ref{fig:clusters} reveal distinct AG patterns across two clusters but show similar AG patterns from different corpora within the same cluster. This behavior is consistent across all the plots.

\subsubsection{Cross-Corpus AG Cluster Overlap Analysis}
In this analysis, we aim to identify overlap in AG clusters between two corpora by analyzing vowel-specific AG samples. We process these samples through an AG cluster model to determine their cluster IDs and then calculate the percentage of samples that are similarly clustered across the corpus using Equation \ref{eq:sim}. For example, if Corpus 1 has clusters 1 and 2, and Corpus 2 includes clusters 1, 2, and 3, we focus on the common clusters for similarity evaluation. To account for potential discrepancies and ensure accuracy, we use a 25\% threshold to exclude very less common clusters, averaging similarity measures only for clusters exceeding this threshold. Table \ref{tab:vowel_clusters} shows the cross-corpus AG cluster Overlap analyses results.
{\setlength\abovedisplayskip{1.55mm}
\setlength\belowdisplayskip{1.55mm}
\begin{equation}
Sim   = \frac{1}{K_{c, t}} \sum_{i=1}^{K_{c, t}} \min \left( \frac{N_{1,i}}{N_1}, \frac{N_{2,i}}{N_2}\right) \times 100 
\vspace{-0.1cm}
\label{eq:sim}
\end{equation}}
%where \(N_{common\_clusters}\) represents the number of common clusters, \(C_{i,j}\) is the number of samples in the common clusters \(i\) and \(j\), and \(N_{total}\) is the total number of samples.
%where \( K_{c, t} \) represents the number of common clusters between the two corpora after applying threshold. \( N_{1,i} \) and \( N_{2,i} \)  denotes the number of samples in cluster \( i \) in source and the target respectively. The total number of samples in 1st corpus and 2nd corpus is shown by \( N_1 \) and \( N_2 \) respectively.
where, \( K_{c, t} \) represents the number of common clusters between the two corpora after thresholding. \( N_{1,i} \) and \( N_{2,i} \) denote the number of samples in cluster \( i \) for the source and target, respectively, with \( N_1 \) and \( N_2 \) being the total number of samples in the source and target, respectively.

\begin{table}[]
\centering
\caption{Percentage overlap of AG clusters across corpora for six vowels, including average overlap for each emotion-specific cluster.}
\renewcommand{\arraystretch}{1}
\resizebox{0.9\columnwidth}{!}{%
\begin{tabular}{ccccccc}
\toprule\specialrule{\cmidrulewidth}{0pt}{0pt}
\textbf{AG Cluster} & \normalsize{/\textipa{A}/} & \normalsize{/\textipa{@}/} & \normalsize{/\textipa{E}/} & \normalsize{/\textipa{i}/} & \normalsize{/\textipa{\ae}/} & \normalsize{/\textipa{u}/} \\ 
\hline \hline
Cluster\_1  &  75.3  &  2.1    &  60.6  &  71.1    &  78.9   &  0   \\
Cluster\_2  &  70.4  &  79.4  & 3.6   & 79.9  &  75.3   &  75.5    \\
Cluster\_3  &  61.2  &  64.5    &  77.1     &  0    &  76.7   &  10.8   \\
Cluster\_4  &  0     &  0  &  78.2     &  69.3  &  0     &  76.2 \\
Cluster\_5  &  0     &  0    &  0     &  5.7  &  15.5  &  0 \\
Cluster\_6  &  0     &  66.8  &  0     &  0    &  12.3  &  0   \\
Cluster\_7  &  68.4  &  0    &  18.4  &  2.5  &  0     &  74.5 \\
Cluster\_8  &  7.1  &  0    &  0     &  0  &  0     &  0   \\
Cluster\_9  &  0     &  12.9  &  0  &  0    &  0     &  0\\
Cluster\_10 &  0     &  0    &  13.6  &  0    &  7.4    &  0   \\   
\midrule
\multicolumn{7}{c}{\textbf{Average Cluster Overlap}}\\
\midrule
All &  68.8 &  69.3   &  74.5  &  73.4    &  76.6   &  75.0  \\
Neutral &  \textbf{75.3 }&  68.4    &  \textbf{80.2} &  66.7   &  72.5  &  60.3   \\
Happiness &  \textbf{80.1}  &  74.3   &  70.9  &  \textbf{78.2 }  &  68.6   &  61.4   \\
Anger &  \textbf{73.4 } &  70.6   &  66.1  &  \textbf{75.6}   &  71.2  &  59.3   \\
Sadness &  66.2  &  65.7   &  \textbf{68.3 }&  63.8    &  67.9 &  \textbf{72.5 }\\
\specialrule{\cmidrulewidth}{0pt}{0pt}\bottomrule    
\end{tabular}}
\label{tab:vowel_clusters}
\vspace{-0.5cm}
\end{table}

Table \ref{tab:vowel_clusters} shows clustering results in two sections: cluster-specific overlap for each vowel and average overlap across all clusters for emotion-specific analysis.
%Table \ref{tab:vowel_clusters} presents clustering results in two sections: one showing cluster-specific overlap across corpora for each vowel and the other displaying average overlap for each vowel. It also highlights emotion-specific average cluster overlap. 
Higher values indicate greater AG cluster similarity across corpora for the given vowels. From Table \ref{tab:vowel_clusters}, we observe substantial overlap in certain clusters, suggesting there exist shared articulatory patterns, which is promising for cross-corpus analyses. For instance, the vowel /\textipa{A}/ is clustered into five groups (1, 2, 3, 7, 8), with most showing high overlap, indicating strong consistency in gestures for this vowel. We can observe this pattern in other vowels as well.

Table \ref{tab:vowel_clusters} reveals varying levels of emotion-specific overlap between corpora for different vowel phonemes. For example, the vowel /\textipa{A}/ has a high similarity score of 80.1\% for \emph{Happiness}, indicating a significant overlap in AG patterns across corpora in this emotional context. Likewise, /\textipa{i}/ shows higher similarity scores for \emph{Happiness} (78.2\%) and \emph{Anger} (75.6\%), suggesting consistent articulatory patterns for this vowel across datasets in these emotional states. These results are consistent with previous studies \cite{sadoughi2018expressive, upadhyay2023phonetic}.
%When analyzing emotion-specific overlap percent shown in Table \ref{tab:vowel_clusters}, we observe varying levels of commonality between corpora with respect to different vowel phonemes. For instance, the vowel /\textipa{A}/ shows a high similarity score of 80.1 under \emph{Happiness}, indicating a significant overlap of AG patterns between corpora in this emotional context. Similarly, /\textipa{i}/ displays higher similarity scores under \emph{Happiness} (78.2) and \emph{Anger} (75.6), reflecting consistent articulatory patterns for this vowel across the datasets in these emotional states. These findings align with previous studies \cite{sadoughi2018expressive, upadhyay2023phonetic}.

\subsubsection{AG-Acoustic Features Association Analysis}
To assess how well samples grouped by AG clusters align with their corresponding acoustic features, we model an acoustic cluster system similar to the AG clustering approach described in Section \ref{sec:clustering}, using time-series K-means with 10 clusters for consistency and comparability. This gives us two clustering models: one based on AG and the other on acoustic features. First, we cluster the samples using the AG cluster model. Then, for each AG cluster, we extract the acoustic features of the grouped samples and passed them through the acoustic cluster model. The goal is to assess how many acoustic clusters form within each AG cluster to evaluate the association between AG-based and acoustic clustering.

\begin{figure}[t!]
\centering
    \begin{subfigure}{0.45\textwidth}
       \includegraphics[width=\textwidth]{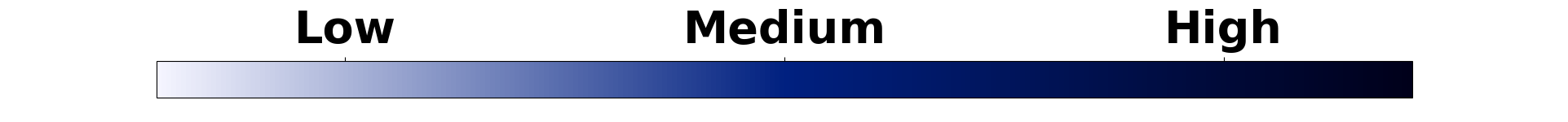}
    \end{subfigure}
    
    \begin{subfigure}{0.15\textwidth}
       \includegraphics[width=\textwidth]{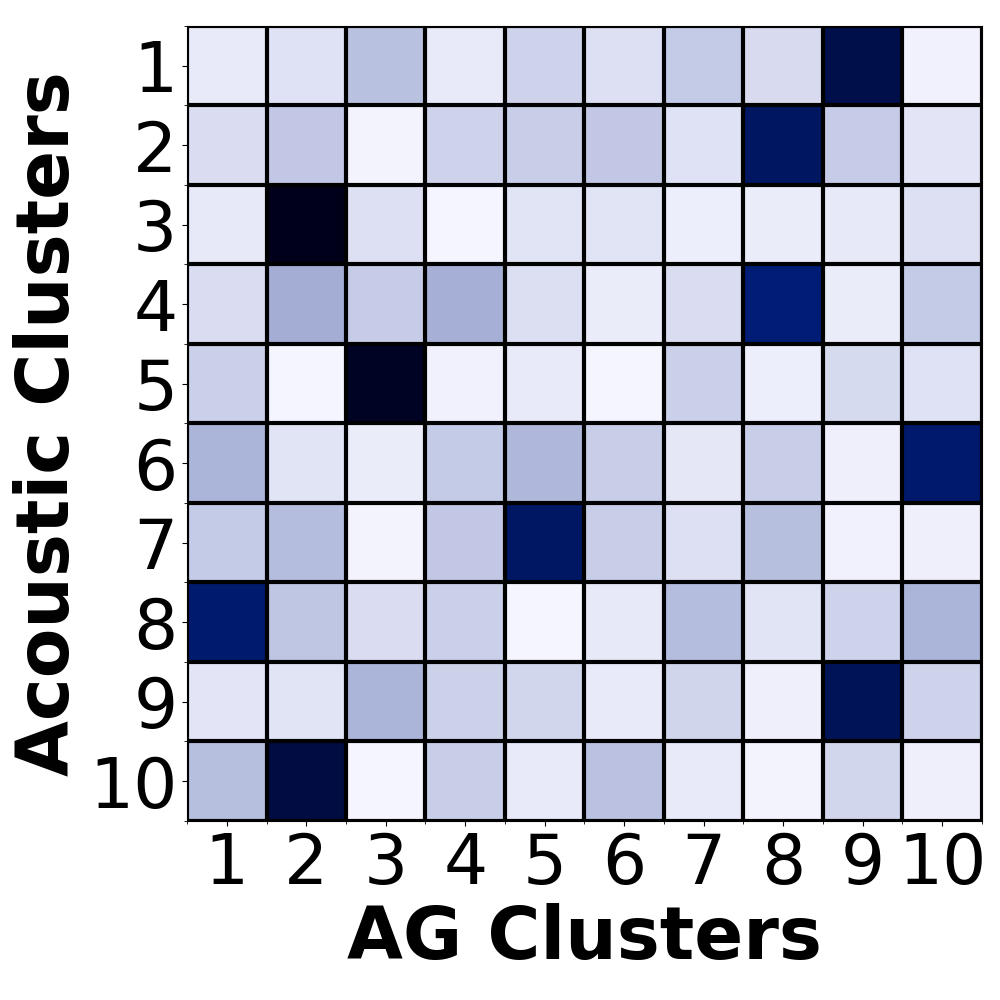}
       \subcaption{\emph{Happiness}}
       \label{fig:hap}
    \end{subfigure}
    \begin{subfigure}{0.15\textwidth}
       \includegraphics[width=\textwidth]{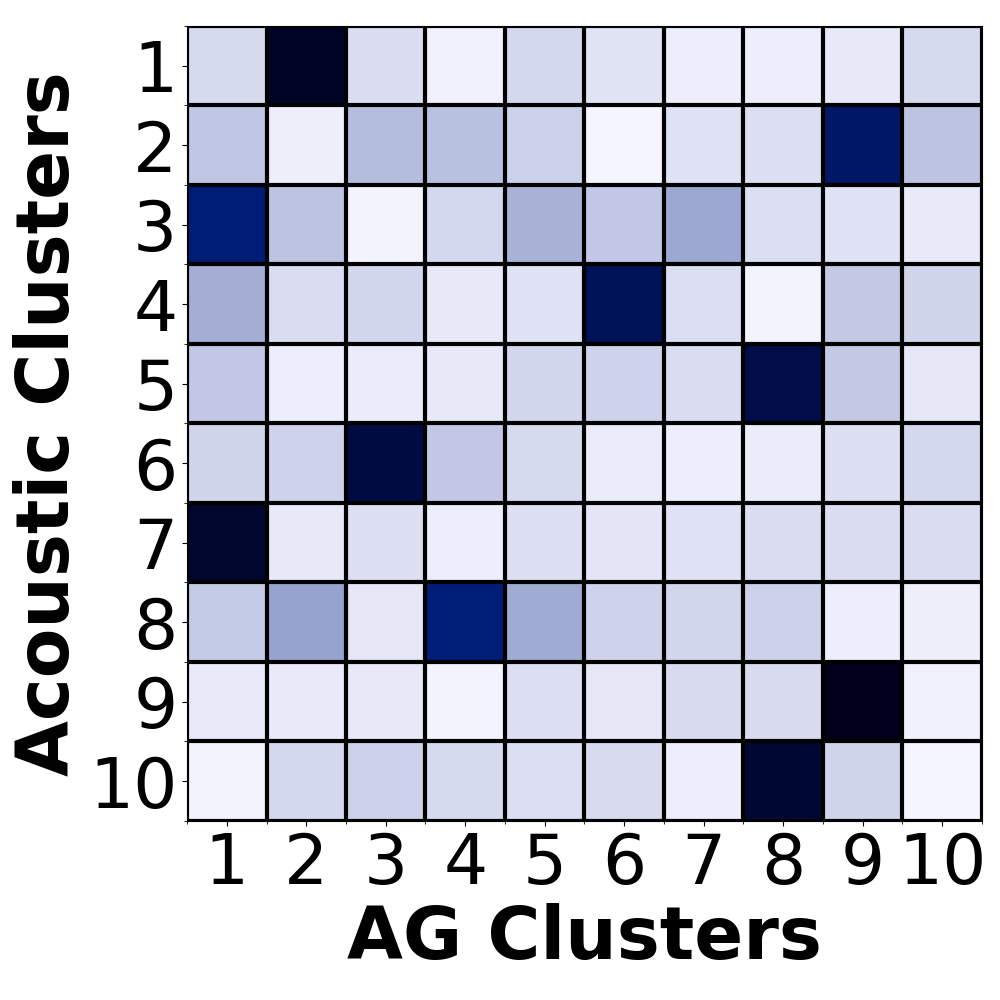}
       \subcaption{\emph{Anger}}
       \label{fig:ang}
    \end{subfigure}   
    \begin{subfigure}{0.15\textwidth}
       \includegraphics[width=\textwidth]{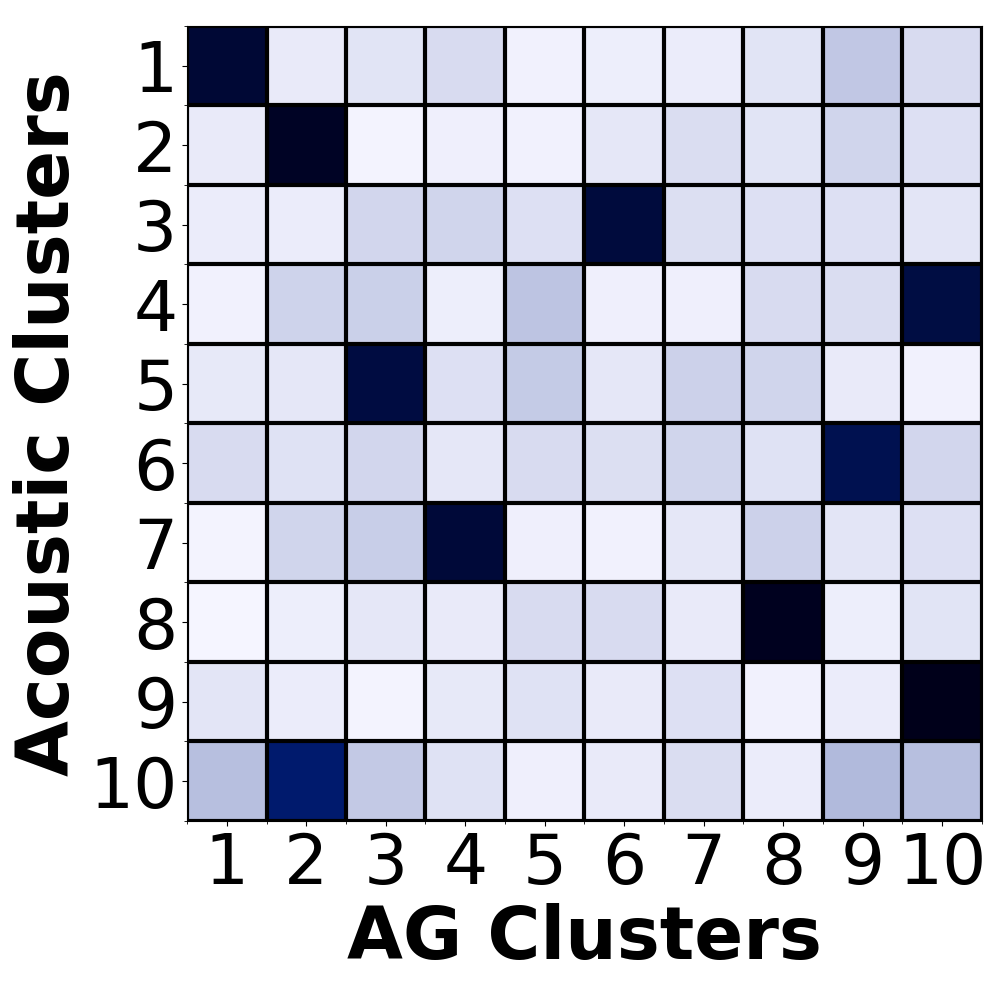}
       \subcaption{\emph{Sadness}}
       \label{fig:sad}
            
    \end{subfigure}  
 \caption{Visualization of association between AG cluster and acoustic cluster across different emotions.}
\label{fig:AG-Acoustic}
\vspace{-0.5cm}
\end{figure}

We visualize the relationship between AG clusters and acoustic clusters using a heatmap, with rows representing AG clusters and columns representing acoustic clusters. Each cell shows how samples from an AG cluster are distributed across acoustic clusters. Given that the AG and acoustic models are trained separately, a diagonal pattern is not expected. However, if samples from an AG cluster mostly align with fewer acoustic clusters, it indicates a stronger association between the AG and acoustic features. Fig. \ref{fig:AG-Acoustic} illustrates that acoustic embeddings often cluster in less number of clusters within each AG cluster, reflecting that there is a correlation between AG gestures and their acoustic counterparts. However, no consistent pattern is observed across acoustic clusters. For example, AG cluster\_1 aligns mostly with acoustic cluster\_9 in \emph{Happiness} but with clusters\_2 in \emph{Anger}. This suggests a one-to-one mapping from AG clusters to acoustic clusters, where each type of mouth articulation corresponds to a specific set of acoustic features. In contrast, the mapping from acoustic clusters to AG clusters is more one-to-many, indicating that a single type of acoustic feature can arise from various mouth articulations.

\section{AG-Anchored Cross-Corpus SER}
\label{sec:proposed}

To improve the cross-corpus emotion transfer task,  we propose an AG clusters-based cross-modal anchoring method for the 4-category SER. Here, we implement a constraint on the common AG clusters over both corpora to align their acoustic space features, called AG-anchored loss \((\mathcal{L}_{\text{AG}}\)) shown in Equation \ref{eq:triplet}. For each target sample, we form triplets using: (1) Anchor: the acoustic embedding of the target sample from the common AG clusters, (2) Positive: an embedding from the same AG cluster and emotion category but from the source dataset, and (3) Negative: an embedding from a different AG cluster within the same emotion category from the source corpus. We adjust the distance between the anchor and negative samples using a weight factor \(w_{i,n}\), based on cluster centroid distances, and compute the soft triplet loss accordingly as shown in Equation \ref{eq:triplet}.

% To enhance cross-corpus emotion transfer, we propose an AG clusters-based cross-modal anchoring method for four-category speech emotion recognition (SER). This method uses the AG-anchored loss \((\mathcal{L}_{\text{AG}}\)) shown in Equation \ref{eq:triplet}, to align acoustic features across corpora by constraining common AG clusters. For each target sample, we form triplets: (1) Anchor: the target sample's acoustic embedding from common AG clusters; (2) Positive: an embedding from the same AG cluster and emotion category in the source dataset; (3) Negative: an embedding from a different AG cluster but within the same emotion category from the source corpus. We adjust the distance between the anchor and negative samples using a weight factor \(w_{i,n}\), based on cluster centroid distances, and compute the soft triplet loss accordingly.
\begin{equation}
\vspace{-0.1cm}
\mathcal{L}_{\text{AG}} = \sum_{i} \left[ d(\mathbf{z}_i, \mathbf{z}_p) - w_{i,n} \cdot d(\mathbf{z}_i, \mathbf{z}_n) + \alpha \right]
%_{+}
%\mathcal{L}_{\text{triplet}}^{\text{soft}}
\label{eq:triplet}
\end{equation}
where \(\mathbf{z}_i\), \(\mathbf{z}_p\), and \(\mathbf{z}_n\) represent the anchor, positive, and negative sets, respectively. \(w_{i,n}\) is soft weight, estimate using the Equation \ref{eq:weight}. \(\alpha\) is a margin parameter that enforces a minimum separation between the positive and negative pairs. Here the value of \(\alpha\) is set to a constant value of 0.3.
%The loss function ensures that the anchor-positive distance \(d(\mathbf{z}_i, \mathbf{z}_p)\) is smaller than the anchor-negative distance , 
\begin{equation}
\vspace{-0.1cm}
w_{i,n} =exp(- \beta \cdot d(\mathcal{C}_{k_i}, \mathcal{C}_{k_n}))
\label{eq:weight}
\end{equation}
%where the function \(d(\cdot)\) represents the distance metric, Euclidean distance. The term \(\alpha\) is a margin parameter that enforces a minimum separation between the positive and negative pairs. The notation \([\cdot]_+\) ensures that only positive values contribute to the loss, meaning that if the expression inside the brackets is negative, the loss for that particular triplet is set to zero. 
where \(d(\mathcal{C}_{k_i}, \mathcal{C}_{k_n})\) is the distance between the centroids of the clusters. \(\beta\) is a scaling parameter that controls the influence of cluster distances. Here the \(\beta\) value is set to 0.2.
%We explore \(\beta\) values ranging from 0.1 to 1.0 and find that \(\beta\)=0.2 gives the best results.

\begin{figure}[tbp]
  \centering
  \includegraphics[height=0.5\linewidth,width=\linewidth]{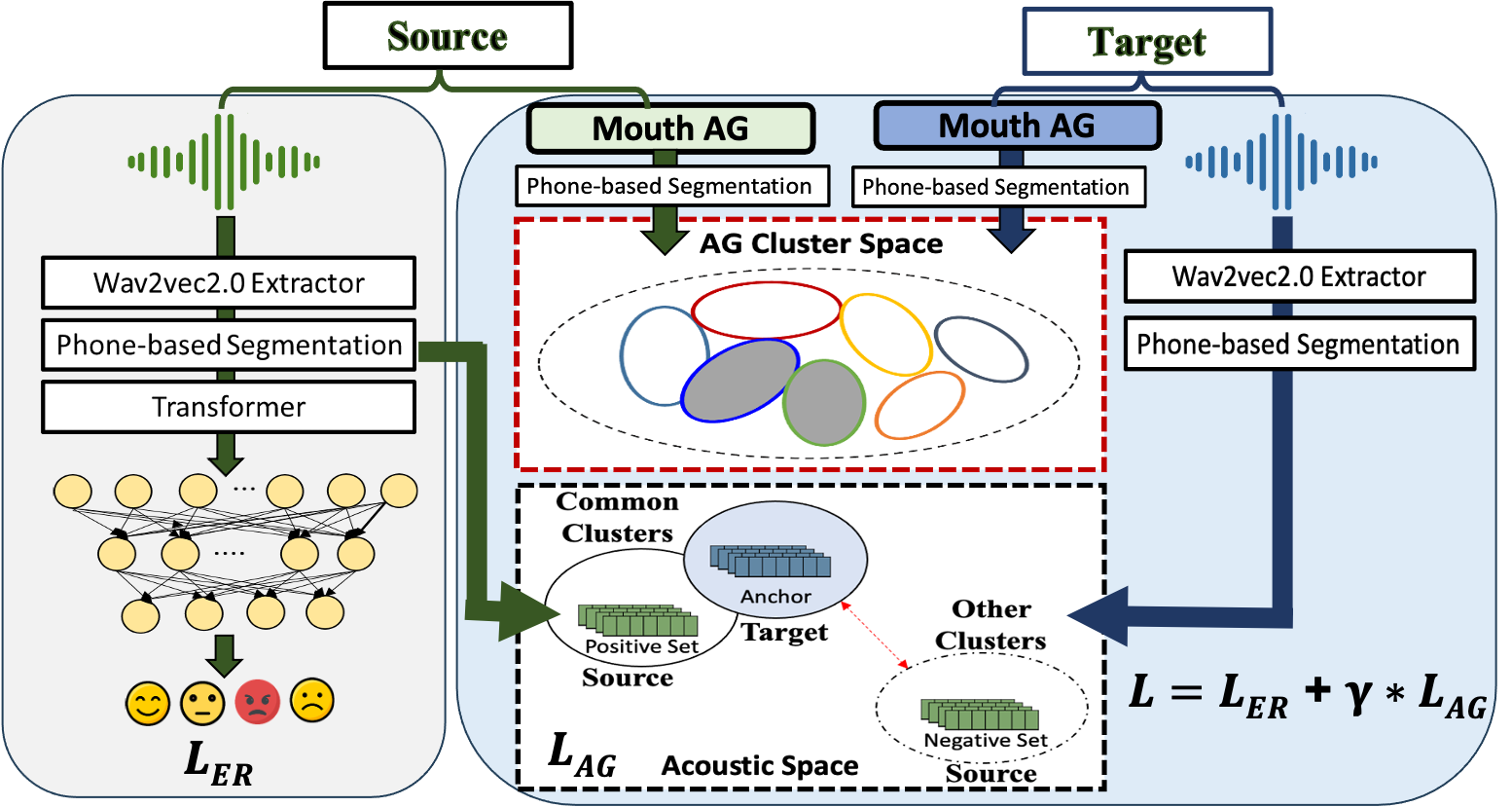}
  \caption{Proposed mouth articulation-based anchoring architecture for cross-corpus SER.}
  \label{fig:arch}
\vspace{-0.2cm}
\end{figure}

This loss function, detailed in Equation \ref{eq:triplet}, ensures that acoustic embeddings from the common AG cluster are closely aligned, while embeddings from different clusters are separated. By incorporating this loss into cross-corpus SER model training, we enhance the alignment of acoustic embeddings based on AG, improving cross-corpus SER. The total loss for each training batch combines the SER loss with the AG-anchored loss, as shown in Equation \ref{eq:total}.
\begin{equation}
\vspace{-0.1cm}
\mathcal{L}_{\text{Total}} =\mathcal{L}_{\text{ER}} + \gamma * \mathcal{L}_{\text{AG}}
\label{eq:total}
\end{equation}
where \(\mathcal{L}_{\text{ER}}\) is the conventional cross-entropy loss for 4-category emotion recognition and the \(\mathcal{L}_{\text{AG}}\) is the soft weighted AG-anchored loss.

\section{Experiment Results}
In our experiments, we benchmark using the CREMA and IMPROV corpora. We utilize Wav2vec2.0 \cite{baevski2020wav2vec} embeddings as pretrained features and apply a transformer with a 4-layer fully connected architecture, similar to our previous work \cite{upadhyay2023phonetic}, Back-propagated using the loss function in Equation \ref{eq:total}, with soft-weighted AG anchoring. The model is optimized with Adam, using a learning rate of 0.0001 and a decay factor of 0.001, and trained for up to 70 epochs with a batch size of 64 with early stopping. The performance is evaluated using the unweighted average recall (UAR) metric.

In our study, we evaluate the effectiveness of the AG-CC method by comparing it with two baseline models: phoneme-anchored (PA-CC) \cite{upadhyay2023phonetic} and layer-anchored (LA-CC) \cite{upadhyay2024layer}. The PA-CC model leverages vowel phonemes as references to align emotional acoustic features in cross-corpora tasks, while the LA-CC approach aligns model layers to maintain consistent emotional patterns across corpora. In contrast, our AG-CC approach aligns these acoustic features through AG clusters to enhance cross-corpus SER. 

The cross-corpus SER performance of all considered models is presented in Table \ref{tab:performance}. As evident from the results, our AG-CC approach outperforms all other models in both the 4-category (4-CAT) and binary SER tasks. Specifically, in the 4-category C→I task, where CREMA is the source and IMPROV is the target, AG-CC achieves a notable performance improvement of 2.02\% over PA-CC and 1.33\% over LA-CC. For binary SER in the C→I tasks, AG-CC delivers strong results across all emotion categories, for instance, \emph{Anger} achieves 75.03\% and \emph{Happiness} reaches 76.74\%. As a sanity check for our model, we also test for the I→C settings, where IMPROV is the source and CREMA is the target. Similar performance patterns are observed, with AG-CC surpassing PA-CC by 0.65\% and LA-CC by 0.88\%. Upper bound results for C→C and I→I SER tasks are also shown in Table \ref{tab:performance}.

To further validate AG-CC, we compare it against the hard-segmented AG anchoring method (Hard-AG), as outline in Table \ref{tab:performance}. In this comparison, Hard-AG refers to the use of fixed AG phonetic segments as anchors, consistent with our previous work \cite{upadhyay2023phonetic} where we anchored on acoustic hard segments. The results, with 54.87\% for C→I and 52.90\% for I→C in the 4-CAT task, indicate that AG-CC’s continuous clustering approach significantly outperforms the Hard-AG method. This is likely due to the continuous nature of AG, which lacks distinct boundaries, making clustering a better fit for capturing their continuous patterns.

%The cross-corpus SER performances of all the considered models are shown in the Table \ref{tab:performance}.Observing Table \ref{tab:performance}, it is evident that our AG-CC approach surpasses all other models in both the 4-category (4-CAT) and binary Speech Emotion Recognition (SER) tasks. Specifically, AG-CC demonstrates a significant performance improvement of 2.02\% over the PA-CC and 1.33\% over the LA-CC in the 4-category C→I task where source is the CREMA and target is the IMPROV. In the binary SER C→I tasks, AG-CC achieves competitive results  over all emotion category, notably reaching 75.03\% for Anger and 76.74\% for Happiness. for  sanity check  of our model we also cross the corpora in I→C SER task where IMPROV is source and CREMA is the target corpora, he we find the similar pattern  as C→I. for instance,  AG-CC shows better performance with increment of 0.65\% and 0.88\% over PA-CC and LA-CC, respectively.   To further validate our idea AG-CC, we compare it against the hard-segmented AG anchoring method (Hard-AG), as shown in Table \ref{tab:performance}. Here  the Hard-AG means the common hard cut  AG phonetic segments are selected as an anchor as we have in our previous work \cite{upadhyay2023phonetic} with acoustic hard segmenst. The result with 54.87 for C→I and 52.90 for I→C over 4-CAT task, indicate that AG-CC's continuous clustering approach outperforms the Haard-AG method. this could be because mouth gesture is hard ....

%\subsection{Performance Comparison}

\begin{table}[t]
\centering
\caption{Cross-corpus SER performance (UAR) for baseline and proposed models: C→I (CREMA to IMPROV) and I→C (IMPROV to CREMA).}
\renewcommand{\arraystretch}{1.2}
\resizebox{0.95\columnwidth}{!}{%
\begin{tabular}{c|c|c|cccc}
\toprule\specialrule{\cmidrulewidth}{0pt}{0pt}
 & & 4-CAT & Neu & Ang & Hap & Sad \\ \hline \hline
\multirow{2}{*}{Upper Bound} & C→C & 66.36  & 89.44 & 88.27  &85.35  & 79.51\\
 & I→I  & 62.10 & 87.84 & 85.33  &83.68 & 75.05  \\\hline 
\multirow{2}{*}{\footnotesize{PA-CC  \cite{upadhyay2023phonetic}}} & C→I & 55.33 & 75.35 & 73.33  &74.82  & 66.98  \\
 & I→C &53.18  & 75.46 & 72.14 & 73.64 & 63.35 \\\hline
\multirow{2}{*}{\footnotesize{LA-CC \cite{upadhyay2024layer}}} & C→I & 56.04 & 78.24  &75.49  &73.50  &66.73  \\
 & I→C & 52.95  &\textbf{ 77.67}  &76.52  & 74.52 &\textbf{67.73}  \\
 \toprule\specialrule{\cmidrulewidth}{0pt}{0pt}
\multirow{2}{*}{\footnotesize{\textbf{AG-CC}}} & C→I & \textbf{57.37} & \textbf{78.51 }& \textbf{75.03}  &\textbf{76.74}  &\textbf{68.10} \\
 & I→C &\textbf{53.83} & 77.46 & \textbf{77.72} & \textbf{75.30} & 65.85\\
\toprule\specialrule{\cmidrulewidth}{0pt}{0pt}
\multirow{2}{*}{\footnotesize{Hard-AG}} & C→I & 54.87 &72.35  &74.46  &71.90  & 69.41 \\
 & I→C &52.90 &70.68  & 71.53 & 70.24  &63.24  \\\hline
%\multirow{2}{*}{\footnotesize{NW-AG} }& & & &  &  \\
 %&& &  &  & \\\hline 
\specialrule{\cmidrulewidth}{0pt}{0pt}\bottomrule                        
\end{tabular}}
\label{tab:performance}
\vspace{-0.1cm}
\end{table}

%\textcolor{red}{write the limitations}

\section{Conclusion}

We introduce the mouth articulation-based anchoring (AG-CC) approach to improve cross-corpus SER by aligning acoustic features across corpora through stable \emph{articulatory gesture} (AG). By focusing on AG, which is more stable than acoustic features, we aim to enhance the generalization of SER systems across different domain corpora. The AG-CC method leverages stable AG anchors for cross-modal alignment, offering a robust foundation for emotion transfer. Our model AG-CC shows the better UAR with 57.37\% for the 4-category cross-corpus SER task. Future work will concentrate on applying this concept in cross-lingual settings, improving AG-clustering strategies to better handle speaker variability, extending AG-CC evaluation to additional emotional categories, and optimizing performance across diverse acoustic environments.

\newpage

\bibliographystyle{IEEEtran}
\bibliography{refs}

\end{document}